\pdfoutput=1
\documentclass[longauth,twocolumn]{aa}
\usepackage[varg]{txfonts}
\usepackage[pdftex]{graphicx}
\usepackage{wasysym}
\usepackage{url, hyperref}
\usepackage{footmisc}
\usepackage{amssymb}
\usepackage{subfigure}
\usepackage{makecell}
\usepackage{multirow}
\usepackage{natbib}
\usepackage[normalem]{ulem}
\usepackage[export]{adjustbox}
\RequirePackage[switch]{lineno}

\newcommand{\fermi}{$Fermi\mathrm{-LAT}$}
\newcommand{\hess}{H.E.S.S.}
\newcommand{\EC}{$\eta$~Car}
\newcommand{\g}{$\gamma$}

\newcommand{\HMS}[3]{$#1^{\mathrm{h}}#2^{\mathrm{m}}#3^{\mathrm{s}}$}
\newcommand{\DMS}[3]{$#1^\circ #2' #3''$}

\def\lsim{\mathrel{\rlap{\lower4pt\hbox{\hskip1pt$\sim$}}
    \raise1pt\hbox{$<$}}}                
\def\gsim{\mathrel{\rlap{\lower4pt\hbox{\hskip1pt$\sim$}}
    \raise1pt\hbox{$>$}}}                

\begin{document}

\title{Detection of very-high-energy \g-ray emission from
  the colliding wind binary \EC\ with \hess}

\titlerunning{Detection of \g-rays from \EC\ with \hess}
\authorrunning{\hess\ Collaboration}

\author{H.E.S.S. Collaboration
\and H.~Abdalla \inst{\ref{NWU}}
\and R.~Adam \inst{\ref{LLR}}
\and F.~Aharonian \inst{\ref{MPIK},\ref{DIAS},\ref{RAU}}
\and F.~Ait~Benkhali \inst{\ref{MPIK}}
\and E.O.~Ang\"uner \inst{\ref{CPPM}}
\and M.~Arakawa \inst{\ref{Rikkyo}}
\and C.~Arcaro \inst{\ref{NWU}}
\and C.~Armand \inst{\ref{LAPP}}
\and T.~Armstrong \inst{\ref{Oxford}}
\and H.~Ashkar \inst{\ref{IRFU}}
\and M.~Backes \inst{\ref{UNAM},\ref{NWU}}
\and V.~Barbosa~Martins \inst{\ref{DESY}}
\and M.~Barnard \inst{\ref{NWU}}
\and Y.~Becherini \inst{\ref{Linnaeus}}
\and D.~Berge \inst{\ref{DESY}}
\and K.~Bernl\"ohr \inst{\ref{MPIK}}
\and R.~Blackwell \inst{\ref{Adelaide}}
\and M.~B\"ottcher \inst{\ref{NWU}}
\and C.~Boisson \inst{\ref{LUTH}}
\and J.~Bolmont \inst{\ref{LPNHE}}
\and S.~Bonnefoy \inst{\ref{DESY}}
\and J.~Bregeon \inst{\ref{LUPM}}
\and M.~Breuhaus \inst{\ref{MPIK}}
\and F.~Brun \inst{\ref{IRFU}}
\and P.~Brun \inst{\ref{IRFU}}
\and M.~Bryan \inst{\ref{GRAPPA}}
\and M.~B\"{u}chele \inst{\ref{ECAP}}
\and T.~Bulik \inst{\ref{UWarsaw}}
\and T.~Bylund \inst{\ref{Linnaeus}}
\and S.~Caroff \inst{\ref{LPNHE}}
\and A.~Carosi \inst{\ref{LAPP}}
\and S.~Casanova \inst{\ref{IFJPAN},\ref{MPIK}}
\and M.~Cerruti \inst{\ref{LPNHE},\ref{CerrutiNowAt}}
\and T.~Chand \inst{\ref{NWU}}
\and S.~Chandra \inst{\ref{NWU}}
\and A.~Chen \inst{\ref{WITS}}
\and S.~Colafrancesco \inst{\ref{WITS}} \protect\footnotemark[2] 
\and G.~Cotter \inst{\ref{Oxford}}
\and M.~Cury{\l}o \inst{\ref{UWarsaw}}
\and I.D.~Davids \inst{\ref{UNAM}}
\and J.~Davies \inst{\ref{Oxford}}
\and C.~Deil \inst{\ref{MPIK}}
\and J.~Devin \inst{\ref{CENBG}}
\and P.~deWilt \inst{\ref{Adelaide}}
\and L.~Dirson \inst{\ref{HH}}
\and A.~Djannati-Ata\"i \inst{\ref{APC}}
\and A.~Dmytriiev \inst{\ref{LUTH}}
\and A.~Donath \inst{\ref{MPIK}}
\and V.~Doroshenko \inst{\ref{IAAT}}
\and J.~Dyks \inst{\ref{NCAC}}
\and K.~Egberts \inst{\ref{UP}}
\and F.~Eichhorn \inst{\ref{ECAP}}
\and G.~Emery \inst{\ref{LPNHE}}
\and J.-P.~Ernenwein \inst{\ref{CPPM}}
\and S.~Eschbach \inst{\ref{ECAP}}
\and K.~Feijen \inst{\ref{Adelaide}}
\and S.~Fegan \inst{\ref{LLR}}
\and A.~Fiasson \inst{\ref{LAPP}}
\and G.~Fontaine \inst{\ref{LLR}}
\and S.~Funk \inst{\ref{ECAP}}
\and M.~F\"u{\ss}ling\footnotemark[1] \inst{\ref{DESY}}
\and S.~Gabici \inst{\ref{APC}}
\and Y.A.~Gallant \inst{\ref{LUPM}}
\and F.~Gat{\'e} \inst{\ref{LAPP}}
\and G.~Giavitto \inst{\ref{DESY}}
\and L.~Giunti \inst{\ref{APC}}
\and D.~Glawion \inst{\ref{LSW}}
\and J.F.~Glicenstein \inst{\ref{IRFU}}
\and D.~Gottschall \inst{\ref{IAAT}}
\and M.-H.~Grondin \inst{\ref{CENBG}}
\and J.~Hahn \inst{\ref{MPIK}}
\and M.~Haupt \inst{\ref{DESY}}
\and G.~Heinzelmann \inst{\ref{HH}}
\and G.~Henri \inst{\ref{Grenoble}}
\and G.~Hermann \inst{\ref{MPIK}}
\and J.A.~Hinton \inst{\ref{MPIK}}
\and W.~Hofmann \inst{\ref{MPIK}}
\and C.~Hoischen \inst{\ref{UP}}
\and T.~L.~Holch \inst{\ref{HUB}}
\and M.~Holler \inst{\ref{LFUI}}
\and M.~H\"{o}rbe \inst{\ref{Oxford}}
\and D.~Horns \inst{\ref{HH}}
\and D.~Huber \inst{\ref{LFUI}}
\and H.~Iwasaki \inst{\ref{Rikkyo}}
\and M.~Jamrozy \inst{\ref{UJK}}
\and D.~Jankowsky \inst{\ref{ECAP}}
\and F.~Jankowsky \inst{\ref{LSW}}
\and A.~Jardin-Blicq \inst{\ref{MPIK}}
\and V.~Joshi \inst{\ref{ECAP}}
\and I.~Jung-Richardt \inst{\ref{ECAP}}
\and M.A.~Kastendieck \inst{\ref{HH}}
\and K.~Katarzy{\'n}ski \inst{\ref{NCUT}}
\and M.~Katsuragawa \inst{\ref{KAVLI}}
\and U.~Katz \inst{\ref{ECAP}}
\and D.~Khangulyan \inst{\ref{Rikkyo}}
\and B.~Kh\'elifi \inst{\ref{APC}}
\and J.~King \inst{\ref{LSW}}
\and S.~Klepser \inst{\ref{DESY}}
\and W.~Klu\'{z}niak \inst{\ref{NCAC}}
\and Nu.~Komin \inst{\ref{WITS}}
\and K.~Kosack \inst{\ref{IRFU}}
\and D.~Kostunin \inst{\ref{DESY}} 
\and M.~Kreter \inst{\ref{NWU}}
\and G.~Lamanna \inst{\ref{LAPP}}
\and A.~Lemi\`ere \inst{\ref{APC}}
\and M.~Lemoine-Goumard \inst{\ref{CENBG}}
\and J.-P.~Lenain \inst{\ref{LPNHE}}
\and E.~Leser\footnotemark[1] \inst{\ref{UP},\ref{DESY}}
\and C.~Levy \inst{\ref{LPNHE}}
\and T.~Lohse \inst{\ref{HUB}}
\and I.~Lypova \inst{\ref{DESY}}
\and J.~Mackey \inst{\ref{DIAS}}
\and J.~Majumdar \inst{\ref{DESY}}
\and D.~Malyshev \inst{\ref{IAAT}}
\and D.~Malyshev \inst{\ref{ECAP}}
\and V.~Marandon \inst{\ref{MPIK}}
\and P.~Marchegiani \inst{\ref{WITS}}
\and A.~Marcowith \inst{\ref{LUPM}}
\and A.~Mares \inst{\ref{CENBG}}
\and G.~Mart\'i-Devesa \inst{\ref{LFUI}}
\and R.~Marx \inst{\ref{MPIK}}
\and G.~Maurin \inst{\ref{LAPP}}
\and P.J.~Meintjes \inst{\ref{UFS}}
\and R.~Moderski \inst{\ref{NCAC}}
\and M.~Mohamed \inst{\ref{LSW}}
\and L.~Mohrmann \inst{\ref{ECAP}}
\and C.~Moore \inst{\ref{Leicester}}
\and P.~Morris \inst{\ref{Oxford}}
\and E.~Moulin \inst{\ref{IRFU}}
\and J.~Muller \inst{\ref{LLR}}
\and T.~Murach \inst{\ref{DESY}}
\and S.~Nakashima  \inst{\ref{RIKKEN}}
\and K.~Nakashima \inst{\ref{ECAP}}
\and M.~de~Naurois \inst{\ref{LLR}}
\and H.~Ndiyavala  \inst{\ref{NWU}}
\and F.~Niederwanger \inst{\ref{LFUI}}
\and J.~Niemiec \inst{\ref{IFJPAN}}
\and L.~Oakes \inst{\ref{HUB}}
\and P.~O'Brien \inst{\ref{Leicester}}
\and H.~Odaka \inst{\ref{Tokyo}}
\and S.~Ohm\footnotemark[1] \inst{\ref{DESY}}
\and E.~de~Ona~Wilhelmi \inst{\ref{DESY}}
\and M.~Ostrowski \inst{\ref{UJK}}
\and M.~Panter \inst{\ref{MPIK}}
\and R.D.~Parsons \inst{\ref{MPIK}}
\and C.~Perennes \inst{\ref{LPNHE}}
\and P.-O.~Petrucci \inst{\ref{Grenoble}}
\and B.~Peyaud \inst{\ref{IRFU}}
\and Q.~Piel \inst{\ref{LAPP}}
\and S.~Pita \inst{\ref{APC}}
\and V.~Poireau \inst{\ref{LAPP}}
\and A.~Priyana~Noel \inst{\ref{UJK}}
\and D.A.~Prokhorov \inst{\ref{WITS}}
\and H.~Prokoph \inst{\ref{DESY}}
\and G.~P\"uhlhofer \inst{\ref{IAAT}}
\and M.~Punch \inst{\ref{APC},\ref{Linnaeus}}
\and A.~Quirrenbach \inst{\ref{LSW}}
\and S.~Raab \inst{\ref{ECAP}}
\and R.~Rauth \inst{\ref{LFUI}}
\and A.~Reimer \inst{\ref{LFUI}}
\and O.~Reimer \inst{\ref{LFUI}}
\and Q.~Remy \inst{\ref{LUPM}}
\and M.~Renaud \inst{\ref{LUPM}}
\and F.~Rieger \inst{\ref{MPIK}}
\and L.~Rinchiuso \inst{\ref{IRFU}}
\and C.~Romoli \inst{\ref{MPIK}}
\and G.~Rowell \inst{\ref{Adelaide}}
\and B.~Rudak \inst{\ref{NCAC}}
\and E.~Ruiz-Velasco \inst{\ref{MPIK}}
\and V.~Sahakian \inst{\ref{YPI}}
\and S.~Sailer \inst{\ref{MPIK}}
\and S.~Saito \inst{\ref{Rikkyo}}
\and D.A.~Sanchez \inst{\ref{LAPP}}
\and A.~Santangelo \inst{\ref{IAAT}}
\and M.~Sasaki \inst{\ref{ECAP}}
\and M.~Scalici \inst{\ref{IAAT}}
\and R.~Schlickeiser \inst{\ref{RUB}}
\and F.~Sch\"ussler \inst{\ref{IRFU}}
\and A.~Schulz \inst{\ref{DESY}}
\and H.M.~Schutte \inst{\ref{NWU}}
\and U.~Schwanke \inst{\ref{HUB}}
\and S.~Schwemmer \inst{\ref{LSW}}
\and M.~Seglar-Arroyo \inst{\ref{IRFU}}
\and M.~Senniappan \inst{\ref{Linnaeus}}
\and A.S.~Seyffert \inst{\ref{NWU}}
\and N.~Shafi \inst{\ref{WITS}}
\and K.~Shiningayamwe \inst{\ref{UNAM}}
\and R.~Simoni \inst{\ref{GRAPPA}}
\and A.~Sinha \inst{\ref{APC}}
\and H.~Sol \inst{\ref{LUTH}}
\and A.~Specovius \inst{\ref{ECAP}}
\and S.~Spencer \inst{\ref{Oxford}}
\and M.~Spir-Jacob \inst{\ref{APC}}
\and {\L.}~Stawarz \inst{\ref{UJK}}
\and R.~Steenkamp \inst{\ref{UNAM}}
\and C.~Stegmann \inst{\ref{UP},\ref{DESY}}
\and C.~Steppa \inst{\ref{UP}}
\and T.~Takahashi  \inst{\ref{KAVLI}}
\and T.~Tavernier \inst{\ref{IRFU}}
\and A.M.~Taylor \inst{\ref{DESY}}
\and R.~Terrier \inst{\ref{APC}}
\and D.~Tiziani \inst{\ref{ECAP}}
\and M.~Tluczykont \inst{\ref{HH}}
\and L.~Tomankova \inst{\ref{ECAP}}
\and C.~Trichard \inst{\ref{LLR}}
\and M.~Tsirou \inst{\ref{LUPM}}
\and N.~Tsuji \inst{\ref{Rikkyo}}
\and R.~Tuffs \inst{\ref{MPIK}}
\and Y.~Uchiyama \inst{\ref{Rikkyo}}
\and D.J.~van~der~Walt \inst{\ref{NWU}}
\and C.~van~Eldik \inst{\ref{ECAP}}
\and C.~van~Rensburg \inst{\ref{NWU}}
\and B.~van~Soelen \inst{\ref{UFS}}
\and G.~Vasileiadis \inst{\ref{LUPM}}
\and J.~Veh \inst{\ref{ECAP}}
\and C.~Venter \inst{\ref{NWU}}
\and P.~Vincent \inst{\ref{LPNHE}}
\and J.~Vink \inst{\ref{GRAPPA}}
\and H.J.~V\"olk \inst{\ref{MPIK}}
\and T.~Vuillaume \inst{\ref{LAPP}}
\and Z.~Wadiasingh \inst{\ref{NWU}}
\and S.J.~Wagner \inst{\ref{LSW}}
\and J.~Watson \inst{\ref{Oxford}}
\and F.~Werner \inst{\ref{MPIK}}
\and R.~White \inst{\ref{MPIK}}
\and A.~Wierzcholska \inst{\ref{IFJPAN},\ref{LSW}}
\and R.~Yang \inst{\ref{MPIK}}
\and H.~Yoneda \inst{\ref{KAVLI}}
\and M.~Zacharias \inst{\ref{NWU}}
\and R.~Zanin \inst{\ref{MPIK}}
\and A.A.~Zdziarski \inst{\ref{NCAC}}
\and A.~Zech \inst{\ref{LUTH}}
\and J.~Zorn \inst{\ref{MPIK}}
\and N.~\.Zywucka \inst{\ref{NWU}}
}

\institute{
Centre for Space Research, North-West University, Potchefstroom 2520, South Africa \label{NWU} \and 
Universit\"at Hamburg, Institut f\"ur Experimentalphysik, Luruper Chaussee 149, D 22761 Hamburg, Germany \label{HH} \and 
Max-Planck-Institut f\"ur Kernphysik, P.O. Box 103980, D 69029 Heidelberg, Germany \label{MPIK} \and 
Dublin Institute for Advanced Studies, 31 Fitzwilliam Place, Dublin 2, Ireland \label{DIAS} \and 
High Energy Astrophysics Laboratory, RAU,  123 Hovsep Emin St  Yerevan 0051, Armenia \label{RAU} \and
Yerevan Physics Institute, 2 Alikhanian Brothers St., 375036 Yerevan, Armenia \label{YPI} \and
Institut f\"ur Physik, Humboldt-Universit\"at zu Berlin, Newtonstr. 15, D 12489 Berlin, Germany \label{HUB} \and
University of Namibia, Department of Physics, Private Bag 13301, Windhoek, Namibia, 12010 \label{UNAM} \and
GRAPPA, Anton Pannekoek Institute for Astronomy, University of Amsterdam,  Science Park 904, 1098 XH Amsterdam, The Netherlands \label{GRAPPA} \and
Department of Physics and Electrical Engineering, Linnaeus University,  351 95 V\"axj\"o, Sweden \label{Linnaeus} \and
Institut f\"ur Theoretische Physik, Lehrstuhl IV: Weltraum und Astrophysik, Ruhr-Universit\"at Bochum, D 44780 Bochum, Germany \label{RUB} \and
Institut f\"ur Astro- und Teilchenphysik, Leopold-Franzens-Universit\"at Innsbruck, A-6020 Innsbruck, Austria \label{LFUI} \and
School of Physical Sciences, University of Adelaide, Adelaide 5005, Australia \label{Adelaide} \and
LUTH, Observatoire de Paris, PSL Research University, CNRS, Universit\'e Paris Diderot, 5 Place Jules Janssen, 92190 Meudon, France \label{LUTH} \and
Sorbonne Universit\'e, Universit\'e Paris Diderot, Sorbonne Paris Cit\'e, CNRS/IN2P3, Laboratoire de Physique Nucl\'eaire et de Hautes Energies, LPNHE, 4 Place Jussieu, F-75252 Paris, France \label{LPNHE} \and
Laboratoire Univers et Particules de Montpellier, Universit\'e Montpellier, CNRS/IN2P3,  CC 72, Place Eug\`ene Bataillon, F-34095 Montpellier Cedex 5, France \label{LUPM} \and
IRFU, CEA, Universit\'e Paris-Saclay, F-91191 Gif-sur-Yvette, France \label{IRFU} \and
Astronomical Observatory, The University of Warsaw, Al. Ujazdowskie 4, 00-478 Warsaw, Poland \label{UWarsaw} \and
Aix Marseille Universit\'e, CNRS/IN2P3, CPPM, Marseille, France \label{CPPM} \and
Instytut Fizyki J\c{a}drowej PAN, ul. Radzikowskiego 152, 31-342 Krak{\'o}w, Poland \label{IFJPAN} \and
School of Physics, University of the Witwatersrand, 1 Jan Smuts Avenue, Braamfontein, Johannesburg, 2050 South Africa \label{WITS} \and
Laboratoire d'Annecy de Physique des Particules, Univ. Grenoble Alpes, Univ. Savoie Mont Blanc, CNRS, LAPP, 74000 Annecy, France \label{LAPP} \and
Landessternwarte, Universit\"at Heidelberg, K\"onigstuhl, D 69117 Heidelberg, Germany \label{LSW} \and
Universit\'e Bordeaux, CNRS/IN2P3, Centre d'\'Etudes Nucl\'eaires de Bordeaux Gradignan, 33175 Gradignan, France \label{CENBG} \and
Institut f\"ur Astronomie und Astrophysik, Universit\"at T\"ubingen, Sand 1, D 72076 T\"ubingen, Germany \label{IAAT} \and
Laboratoire Leprince-Ringuet, École Polytechnique, CNRS, Institut Polytechnique de Paris, F-91128 Palaiseau, France \label{LLR} \and
APC, AstroParticule et Cosmologie, Universit\'{e} Paris Diderot, CNRS/IN2P3, CEA/Irfu, Observatoire de Paris, Sorbonne Paris Cit\'{e}, 10, rue Alice Domon et L\'{e}onie Duquet, 75205 Paris Cedex 13, France \label{APC} \and
Univ. Grenoble Alpes, CNRS, IPAG, F-38000 Grenoble, France \label{Grenoble} \and
Department of Physics and Astronomy, The University of Leicester, University Road, Leicester, LE1 7RH, United Kingdom \label{Leicester} \and
Nicolaus Copernicus Astronomical Center, Polish Academy of Sciences, ul. Bartycka 18, 00-716 Warsaw, Poland \label{NCAC} \and
Institut f\"ur Physik und Astronomie, Universit\"at Potsdam,  Karl-Liebknecht-Strasse 24/25, D 14476 Potsdam, Germany \label{UP} \and
Friedrich-Alexander-Universit\"at Erlangen-N\"urnberg, Erlangen Centre for Astroparticle Physics, Erwin-Rommel-Str. 1, D 91058 Erlangen, Germany \label{ECAP} \and
DESY, D-15738 Zeuthen, Germany \label{DESY} \and
Obserwatorium Astronomiczne, Uniwersytet Jagiello{\'n}ski, ul. Orla 171, 30-244 Krak{\'o}w, Poland \label{UJK} \and
Centre for Astronomy, Faculty of Physics, Astronomy and Informatics, Nicolaus Copernicus University,  Grudziadzka 5, 87-100 Torun, Poland \label{NCUT} \and
Department of Physics, University of the Free State,  PO Box 339, Bloemfontein 9300, South Africa \label{UFS} \and
Department of Physics, Rikkyo University, 3-34-1 Nishi-Ikebukuro, Toshima-ku, Tokyo 171-8501, Japan \label{Rikkyo} \and
Kavli Institute for the Physics and Mathematics of the Universe (WPI), The University of Tokyo Institutes for Advanced Study (UTIAS), The University of Tokyo, 5-1-5 Kashiwa-no-Ha, Kashiwa, Chiba, 277-8583, Japan \label{KAVLI} \and
Department of Physics, The University of Tokyo, 7-3-1 Hongo, Bunkyo-ku, Tokyo 113-0033, Japan \label{Tokyo} \and
RIKEN, 2-1 Hirosawa, Wako, Saitama 351-0198, Japan \label{RIKKEN} \and
University of Oxford, Department of Physics, Denys Wilkinson Building, Keble Road, Oxford OX1 3RH, UK \label{Oxford} \and
Now at Physik Institut, Universit\"at Z\"urich, Winterthurerstrasse 190, CH-8057 Z\"urich, Switzerland \label{MitchellNowAt} \and
Now at Institut de Ci\`{e}ncies del Cosmos (ICC UB), Universitat de Barcelona (IEEC-UB), Mart\'{i} Franqu\`es 1, E08028 Barcelona, Spain \label{CerrutiNowAt} 
}

\offprints{H.E.S.S.~collaboration,
\protect\\\email{\href{mailto:contact.hess@hess-experiment.eu}{contact.hess@hess-experiment.eu}};
\protect\\\protect\footnotemark[1] Corresponding authors
\protect\\\protect\footnotemark[2] Deceased
}

\date{}
\abstract
{}
{
  Colliding wind binary systems have long been suspected to be high-energy
  (HE; 100\,MeV$< E < $100\,GeV) \g-ray emitters. \EC\ is the most prominent
  member of this object class and is confirmed to emit phase-locked HE
  \g\ rays from hundreds of MeV to $\sim$100\,GeV
  energies. This work aims to search for and characterise the
  very-high-energy (VHE; $E >$100\,GeV) \g-ray emission from \EC\
  around the last periastron passage in 2014 with the ground-based
  High Energy Stereoscopic System (\hess).
}
{
  The region around \EC\ was observed with \hess\ between orbital
  phase $p = 0.78 - 1.10$, with a closer sampling at $p\approx 0.95$
  and $p\approx 1.10$ (assuming a period of 2023 days). Optimised hardware settings as well as
  adjustments to the data reduction, reconstruction, and
  signal selection were needed to suppress and take into account the
  strong, extended, and inhomogeneous night sky background (NSB) in the
  \EC\ field of view. Tailored run-wise Monte-Carlo simulations
    (RWS) were required to accurately treat the additional noise from
  NSB photons in the instrument response functions.
}
{
  \hess\ detected VHE \g-ray emission from the direction of \EC\
  shortly before and after the minimum in the X-ray light-curve close
  to periastron. Using the point spread function provided by
  RWS, the reconstructed signal is point-like and the
  spectrum is best described by a power law. The overall flux and 
  spectral index in VHE \g\ rays agree within statistical and
  systematic errors before and after periastron. The \g-ray
    spectrum extends up to at
  least $\sim$400\,GeV. This implies a maximum magnetic field
  in a leptonic scenario in the emission region of 0.5\,Gauss. No
  indication for phase-locked flux variations is detected in the
  \hess\ data. 
}
{}


\maketitle

\section{Introduction}

In 2009, the newly launched {\it AGILE} satellite and the Large
Area Telescope (LAT) onboard the {\it Fermi} satellite detected a bright \g-ray
source, which is coincident with the colliding wind binary (CWB) \EC\
\citep{EtaCar:Agile, FERMI:BSL}. \EC\ is composed
of a luminous blue variable primary star of $\sim$100
$\textrm{M}_\odot$ and an O- or B-type companion of $\sim$30
$\textrm{M}_\odot$, and has been the object of numerous observations over
centuries with ground-based telescopes as well as satellites, such as in
radio, millimetre, infrared, optical, ultraviolet, and X-ray
wavelengths \citep[see e.g.][for a review]{Humphreys2012}. The {\it
  AGILE} and \fermi\ detection, however, was the
first time that high-energy \g-ray emission was seen from a CWB. The
two member stars of \EC\ orbit each other in a very eccentric orbit
($e\sim0.9$) with a period of $\sim$2023 days \citep{Damineli2008}. Since
the \g-ray detection in 2009, \fermi\ continued to monitor \EC\ and has
now covered almost two orbits \citep{Balbo2017}. The observed
\g-ray emission is variable and composed of a low- and a high-energy
component of which the latter extends up to $\sim$300\,GeV and shows
orbit-to-orbit changes in the light-curve
\citep{Reitberger2012, Reitberger2015, Balbo2017}.

A CWB system, such as \EC, is typically composed of two massive stars that
orbit each other and whose stellar winds form a colliding wind region (CWR) at the
locations of ram-pressure balance. The CWR is characterised by a
contact discontinuity and a strong shock on either side of it. The
detection of non-thermal radio emission of
prominent systems, such as WR~140, WR~146, and WR~147, have raised
interest in this source class \citep[see e.g.][and references
therein for a list of particle-accelerating
CWBs]{deBecker2013}. However, \EC\ has not been identified in
non-thermal radio emission yet \citep{Duncan1995}, which could be
explained by a significant suppression due to the Razin effect
\citep[e.g.][]{Falceta2012} or synchrotron self-absorption
\citep[e.g.][]{Gupta2017}. The first theoretical works studying
particle acceleration in CWBs emerged in the 1980s and 1990s
\citep[e.g.][]{Casse1980, Voelk1982, Eichler1993, Romero1999} and were
further extended in the following years \citep[e.g.][]{Muecke02, Benaglia03,
  CWB:Reimer06}. 

Both \EC\ member stars drive strong and dense supersonic stellar
winds. The primary star loses
$\dot{M_1} \approx 2\, \times 10^{-4} \,\textrm{M}_\odot \,
\textrm{yr}^{-1}$ in its 500\,km\,s$^{-1}$ fast wind 
  \citep{EtaCar:Pittard02} \footnote{We note that
\citet{Groh2012} find even higher primary mass-loss rates.} The wind of the
companion star has a mass-loss rate of $\dot{M_2} \approx 2
\,\times\,10^{-5} \,\textrm{M}_\odot\,\textrm{yr}^{-1}$, it is
less dense than the primary wind, but is moving at considerably faster
speeds of $v_2 \approx\,3000\,\textrm{km}\,\textrm{s}^{-1}$
\citep{Hillier2001, EtaCar:Pittard02, EtaCar:Parkin09}. In the CWR,
wind material is shock-heated to $50$\,MK and gives rise to soft X-ray
emission \citep{EtaCar:Damineli96, Corcoran2015}. Charged
particles are predicted to be accelerated in the shocks on the primary and
  companion side of the contact discontinuity, reaching energies
  several orders of magnitude higher than those of the shock-heated
  plasma \citep[see e.g.][and references therein]{CWB:Reimer06}.

The \g-ray spectrum as measured with \fermi\ exhibits two spectral
components, one at low and one at high \g-ray energies. The lower-energy
component shows a peak of the emission at $\sim$1\,GeV, followed by a
cutoff.
A second, harder spectral component is visible in the 10\,GeV $-$ 300\,GeV range
\citep{Farnier2011, Reitberger2012}. Both components show
variability at different levels along the orbit, with a prominent rise
in the 10\,GeV $-$ 300\,GeV flux around periastron
\citep{Reitberger2015}. The observed phase-locked flux variations on month-timescales
is nowadays seen as confirmation for the origin of the \EC\ \g-ray
emission in the binary system. The origin of the emission of the
lower-energy component is discussed in the literature in different
frameworks. While \citet{Farnier2011, Bednarek2011, Balbo2017, Hamaguchi2018} favour a leptonic
origin, \citet{Ohm2015} suggest a hadronic interpretation, where
  the emission stems from the decay of neutral pions that were
  produced in hadronic interactions. The high-energy
component of the emission has been suggested to originate
from hadrons \citep[e.g.][]{Farnier2011, Ohm2015,
  Reitberger2015, Balbo2017}. In the CWR, 
electrons suffer strong inverse Compton and synchrotron losses, which
makes acceleration to VHEs challenging and may limit the
maximum energy electrons can be accelerated to. Above energies of 100$\,$GeV
phase-locked flux variations due to absorption,
modulated by orbital motion, are expected
  \citep[e.g.][]{Ohm2015}. Measuring temporal variability of the
emission is key to constrain the \g-ray emission region.

\hess\ observed \EC\ from 2004 to 2010 to search for VHE \g-ray
emission from the binary and the Carina Nebula with the \hess\ phase-I
telescopes \citep{Abramowski2012}, but could only provide upper
limits. Combined with the \fermi\ results, the
\hess\ non-detection of VHE \g-ray emission above $\sim$500\,GeV
  energies implies a cut-off of the accelerated particle
population, or severe losses due to \g-\g\ absorption in the strong
stellar radiation fields. The original \hess\ array was expanded by
installing the CT5 telescope in 2012, which lowered the instrument's
energy threshold significantly. This finally allows \hess\ to reach
into the domain of the \EC\ \g-ray emission. Although detected in high-energy
\g-rays over long exposures, one major limitation of \fermi\ is its
comparably small detection area, which limits the sensitivity to study
short-timescale changes in the \EC\ light-curve. For instance, over a
$\sim$6-month period, the X-ray light-curve \citep{Corcoran2017} shows
a strong increase in flux shortly before periastron, followed by an
X-ray flux minimum, which lasts for $30 - 60$ days, and a recovery shortly
thereafter. The high-energy \g-ray light-curves on the other hand only show
significant detections in phase bins of typically 2.5 (5.0) months
duration in the low (high) energy component
\citep{Reitberger2012}. The energy range of \fermi\ at the
highest energies and \hess\ phase-II at the lowest energies cover the
\EC\ \g-ray spectrum in the cut-off region. \hess\ compared to
\fermi\ has a much larger collection area, which makes it very well
suited to study short-timescale changes in the \EC\ light-curve
especially close to periastron. At 100\,GeV energies, the
\hess\ differential sensitivity of a 15-hour observation is
a factor $\sim$50 better than \fermi\ in a 30-day period
\citep{Hoischen2017}, which motivates the \hess\ observations presented here.

The paper is organised as follows: in section \ref{sec:analysis}
we discuss the H.E.S.S. instrument, the \EC\ data set, and the data analysis. The
impact of the strong and inhomogeneous night sky background (NSB) is
also discussed. In section~\ref{sec:discussion} we conclude with
placing the H.E.S.S. detection in the multiwavelength context and
discuss it within the framework of particle acceleration and \g-ray
production in \EC. 


\section{H.E.S.S. data analysis}
\label{sec:analysis}

\subsection{H.E.S.S. data}
\label{sec:nsb_trigger}
The \hess\ experiment is an array of five Imaging
Atmospheric Cherenkov Telescopes (IACTs) located in Namibia. It
is the only IACT system integrating different telescope types
into one array and is sensitive to \g-ray emission from Galactic and
extragalactic sources in the energy range from
$\sim$$30$\,GeV to tens of TeV. The four H.E.S.S.-I telescopes are
equipped with mirrors of 12\,m diameter and have a Davies-Cotton
  mirror geometry. The larger CT5 telescope has a parabolic mirror diameter of 28\,m,
which makes it the largest IACT worldwide. The field of view (FoV) of
the H.E.S.S.-I telescopes is 5$\,^{\circ}$ in diameter, whereas CT5 has a FoV of
$\sim$3.2$\,^{\circ}$ \citep{Bolmont2014}. The addition of CT5 to the
array significantly lowered the energy threshold, allowing for a
broader overlap in energy in the transition region between \hess\ and space-based
instruments like \fermi.

The main challenge in the analysis of \g-ray data from the direction
of \EC\ is the large number of UV photons originating from the Carina
Nebula that cause a significant increase in NSB, causing fake
triggers of the telescope and increasing noise levels in the image
pixels. For instance, the NSB rate in the Galactic plane is
typically around 100\,MHz per pixel in the image. This translates to a mean
number of NSB photons of 1.6 photo-electrons (p.e.) per
pixel. The NSB in the direction of \EC\ is up to ten times higher than
the Galactic average and varies greatly across the FoV. To distinguish
real Cherenkov showers from
NSB-induced events, at least three neighbouring pixels with
$>4.0$\,p.e. above the average NSB are required to trigger the CT5 camera. The
resulting telescope trigger rate is 1.5 kHz, mostly from genuine air
showers. In the case of \EC\ this threshold was applied for
data set I (DS-I; see Tab.~\ref{tab:data}), while CT5 camera HV
adjustments required to increase the pixel trigger thresholds to
4.5\,p.e. for data set~II (DS-II).

\begin{table}
  \centering
  \caption{Properties of the data sets used in this work to
    calculate the spectral points and the light-curve shown in
    Fig.~\ref{fig:Spec} and \ref{fig:LC}, respectively. The time interval
    of the \hess\ observations, the total live time corresponding to
    the individual data sets, along with the covered orbital phase are summarised.}
  \begin{tabular}{@{}ccccc}
    \hline
    Data Set & Observation Time & Phase & live time \\
    (DS) & Modified Julian Date (MJD) & & (hours) \\ \hline
    I & $56439 - 56444$ & 0.78 & 0.9 \\
      & $56666 - 56667$ & 0.89 & 0.8 \\
      & $56741 - 56751$ & 0.93 & 9.4 \\
      & $56775 - 56778$ & 0.95 & 0.9 \\
      & $56797 - 56799$ & 0.96 & 1.8 \\ \hline
   II & $56803 - 56807$ & 0.96 & 3.3 \\
      & $57066 - 57074$ & 1.09 & 8.6 \\
      & $57077 - 57079$ & 1.10 & 4.0 \\ \hline
 all  & $56439 - 57079$ & $0.78 - 1.10$ & 29.7 \\ \hline
  \end{tabular}
  \label{tab:data}
\end{table}

Observations in the direction of \EC\ were
mainly performed in the first six months of 2014 and 2015 (with the
exception of one observation, which was conducted in May/June
2013). The properties of the individual data sets are summarised
in Tab.~\ref{tab:data}. \EC\ was observed with only CT5 for DS-II,
which limits the analysis to monoscopic
data. The data quality of DS-I and DS-II has been checked with
standard tools. To adjust to the high NSB level and the extra pixels
turned off to protect the Cherenkov camera, the requirement on the
maximum relative fraction of deactivated pixels per observation in CT5 has
been relaxed from 5\% to 10\% \citep[cf][]{2017AGNHessII}.

\subsection{H.E.S.S. data analysis}
\label{sec:hess_analysis}
The \EC\ data have been analysed using two software packages with
independent shower simulation, calibration, reconstruction and
\g/hadron separation to cross check the results. Results presented in
this paper were produced with the \textit{model analysis}
\citep{2009DeNaurois}, which is based on a semi-analytical model for
the electromagnetic shower development in the atmosphere to infer the
properties of the primary particle. All results have been successfully
cross-checked in the {\textit{HAP} framework, which employs a
neural-network based shower reconstruction and \g/hadron separation
\citep{Murach2015}.

\begin{figure*}[ht!]
  \begin{minipage}{0.595\textwidth}
    \centering
    \includegraphics[width=\textwidth]{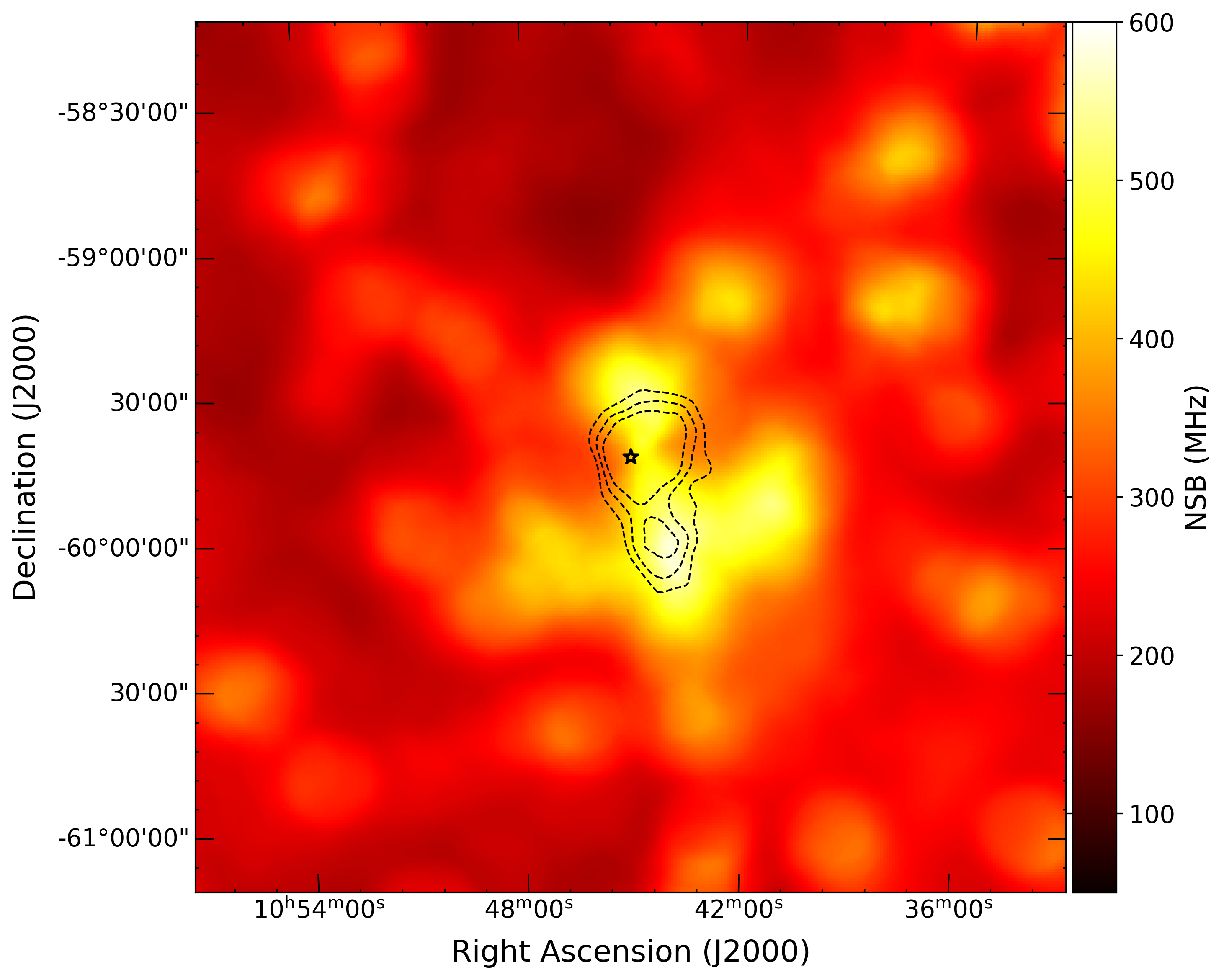}
\end{minipage}
\begin{minipage}{0.395\textwidth}
  \includegraphics[width=0.9\textwidth]{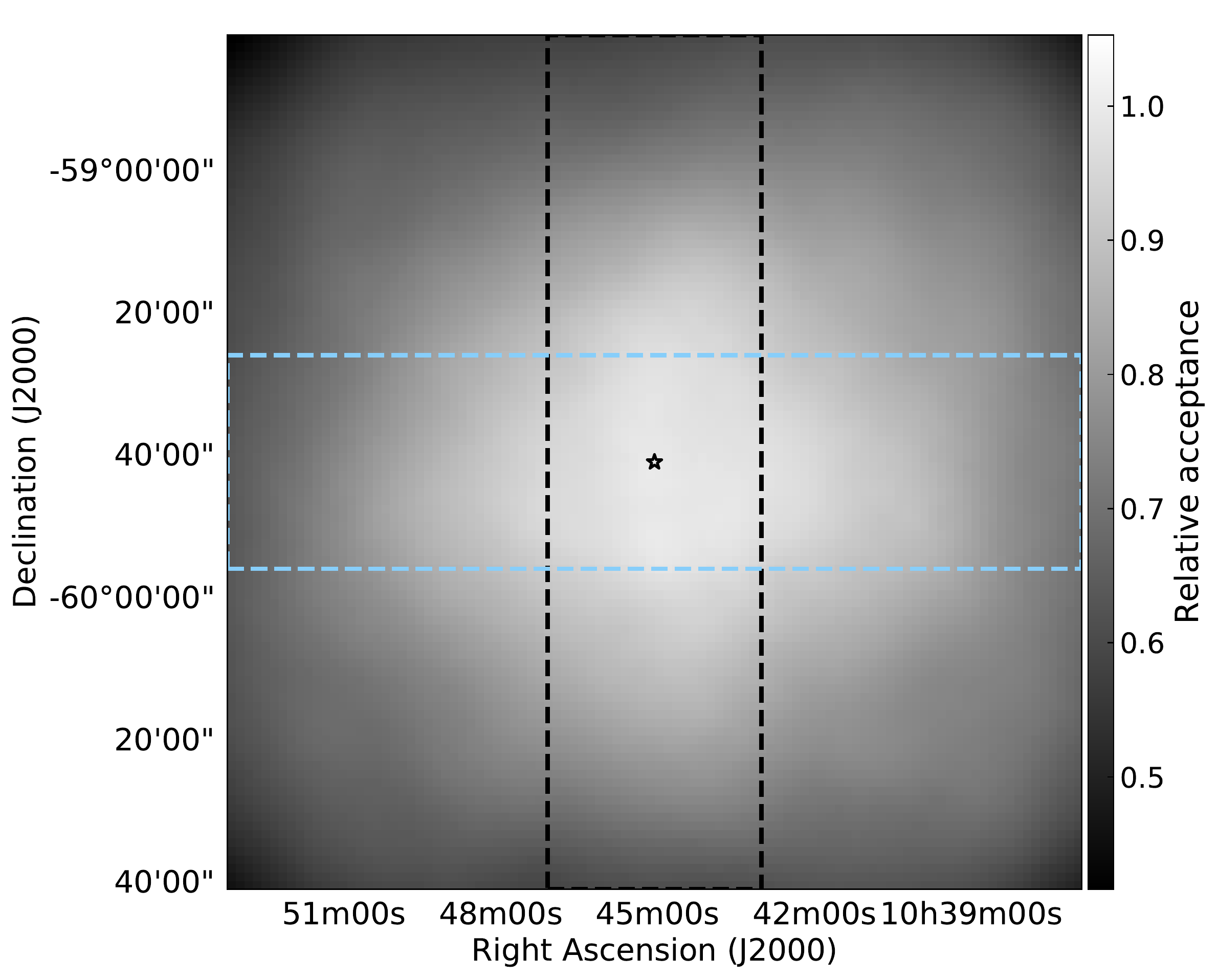}
  \includegraphics[width=\textwidth]{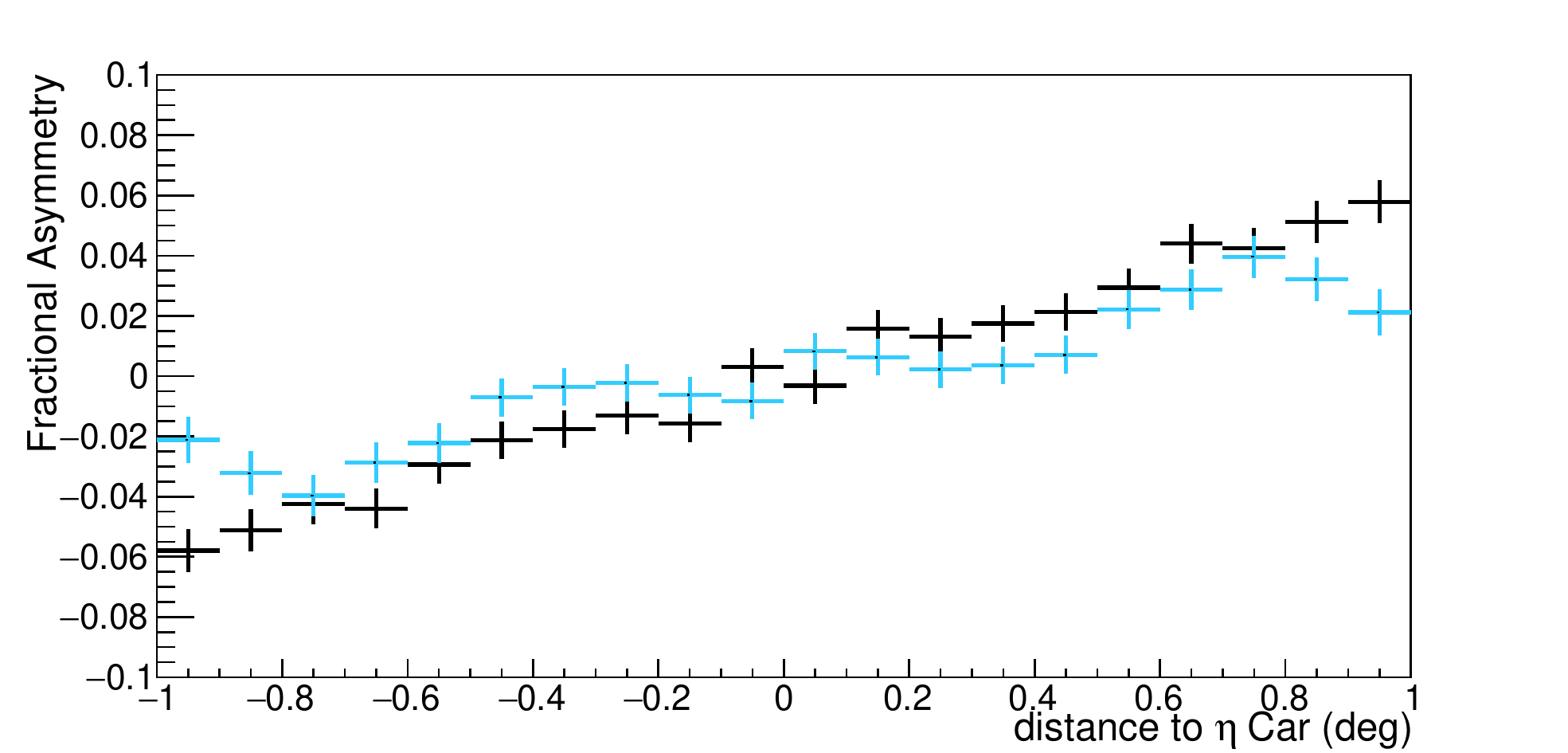}
\end{minipage}
\caption{\label{fig:NSB} {\it Left:} Smoothed NSB rate per camera pixel in MHz
  for CT5 overlaid with contours for 3.5, 4.75 and 6 $\sigma$ of the mono
  analysis of DS-I and DS-II. Turned-off pixels are
    treated as 0\,MHz, hence the map illustrates the average NSB
    operating pixels are exposed to. The optical position of \EC\ is denoted
  with a small star. {\it Right:} Zoom into the 2D \g-ray acceptance 
    map (top) and the relative asymmetry in RA and Dec (bottom).}
\end{figure*}

The NSB in the \EC\ FoV as shown in Fig.~\ref{fig:NSB}
poses serious challenges for the instrument and the \g-ray
data analysis. As described in \citet{Aharonian2004}, the NSB rate can
be inferred for every camera pixel without Cherenkov light signal
based on the pedestal width. Additional checks and systematic tests have been
performed to study its influence on the calibration, shower
reconstruction, and \g/hadron separation. These tests are described in
more detail in the following.

The Carina Nebula is extended over a $\gtrsim0.3^\circ$ region, which is comparable to the Cherenkov image size of low-energy
showers (cf. Fig.~\ref{fig:NSB}). Noise triggers could hence locally
mimic low-energy extensive air showers. The goal
was to reduce the impact of noise-induced triggers as early as
possible in the analysis chain. Therefore, the main and cross-check analysis
apply an image cleaning procedure to provide the seed for the
likelihood fit \citep[see below][]{2009DeNaurois}, and to reject noise pixels in
the image analysis \citet{Aharonian2004}, respectively. The nominal
dual-threshold cleaning of 4\,p.e. for a pixel and 7\,p.e. for a
neighbour pixel (i.e. (4,7) cleaning) has been increased to a (6,12)
cleaning.

As a next step, a pixel-wise log-likelihood comparison between the
predicted Cherenkov shower images, taking into account the NSB and
electronic noise, and the images recorded by the telescopes is
performed. This fit provides an estimate for the energy and direction
of the incident particle \citep{2009DeNaurois, Holler2015}. The
semi-analytical model also provides a \textit{goodness of fit}
parameter and the reconstructed primary interaction depth that are
used to discriminate \g-ray events from hadronic background. In
addition, the estimated error on the reconstructed direction of
each event is required to be smaller than 0.3$^{\circ}$.

To control the NSB, a \textit{uniform NSB goodness} variable was
defined, which characterises the likelihood of accidentally
triggering on fluctuations caused by NSB. In contrast to the
\textit{NSB goodness} variable as introduced in \citet{2017AGNHessII},
the \textit{uniform NSB goodness} parameter assures a flat acceptance
to background-like events, at the cost of an inhomogeneous
acceptance to \g-ray like events. Although the number of accidental
NSB camera triggers are reduced to a manageable level, NSB 
photons are still recorded, when real Cherenkov-induced events
trigger the array and are recorded. The reduction in NSB-only
events by applying these cuts and the response
of the instrument is demonstrated in Fig.~\ref{fig:NSB},
right. Here, the acceptance to \g-ray like background events is
  shown for DS-I and DS-II. The asymmetry in RA and Dec expressed as
  $(x - (-x)) / (x + (-x))$ across the \EC\ FoV is stable at the 5\%$
  - $10\% level. This shows that the
influence of NSB on the background extraction is generally under
control. For the generation of sky images, the \textit{ring
  background} technique is used, while for the spectrum
determination the \textit{reflected background} method is used
\citep{Berge2007}.

\subsubsection{VHE \g-ray detection}
The analysis of data acquired in \hess\ observations towards \EC\
before 2011, and published in \citet{Abramowski2012}, yielded an energy
threshold of 470$\,$GeV. For the CT5 mono analysis presented in
this work, and the modifications made as described above, the
energy threshold is 190\,GeV and 220\,GeV for DS-I and DS-II,
respectively. Figure~\ref{fig:MonoSign} shows the VHE \g-ray
significance map of the sky in the direction of \EC\ for all monoscopic events from 
both data sets that pass \g/hadron separation cuts as described in the
previous section. Significant VHE \g-ray emission is also seen in DS-I
and DS-II separately. In the signal region, centred at the
position of \EC, a \g-ray excess of ($526\pm 62$) events above
background at a significance level of 8.9\,$\sigma$ is found in DS-I. In
DS-II, the \g-ray excess is $541\pm 56$ at a statistical significance
of 10.3\,$\sigma$. The emission seen in DS-I with the mono analysis is
also confirmed, but at a lower significance of 7.2\,$\sigma$, with the
CT1-5 stereo analysis. This can partly be explained by the increased
energy threshold of the stereo analysis. Also, the cross-check analysis
confirms the significant detections in both data sets. 

The maximum of the \g-ray emission shown in Fig.~\ref{fig:MonoSign}
stems from the direction of \EC\ and not from the region of highest
NSB. Outside of the signal region, the distribution of sky map
significance is roughly consistent with the expected normal
distribution, showing a width of 1.04 and 1.08 in DS-I and DS-II,
respectively. The maximum of the NSB, as shown in Fig.~\ref{fig:NSB},
does however coincide with some residual \g-ray like emission south of
\EC. The origin of this feature and its implication for the \EC\
measurement are discussed in the following.

\begin{figure}[ht]
\centering
\includegraphics[width=0.495\textwidth]{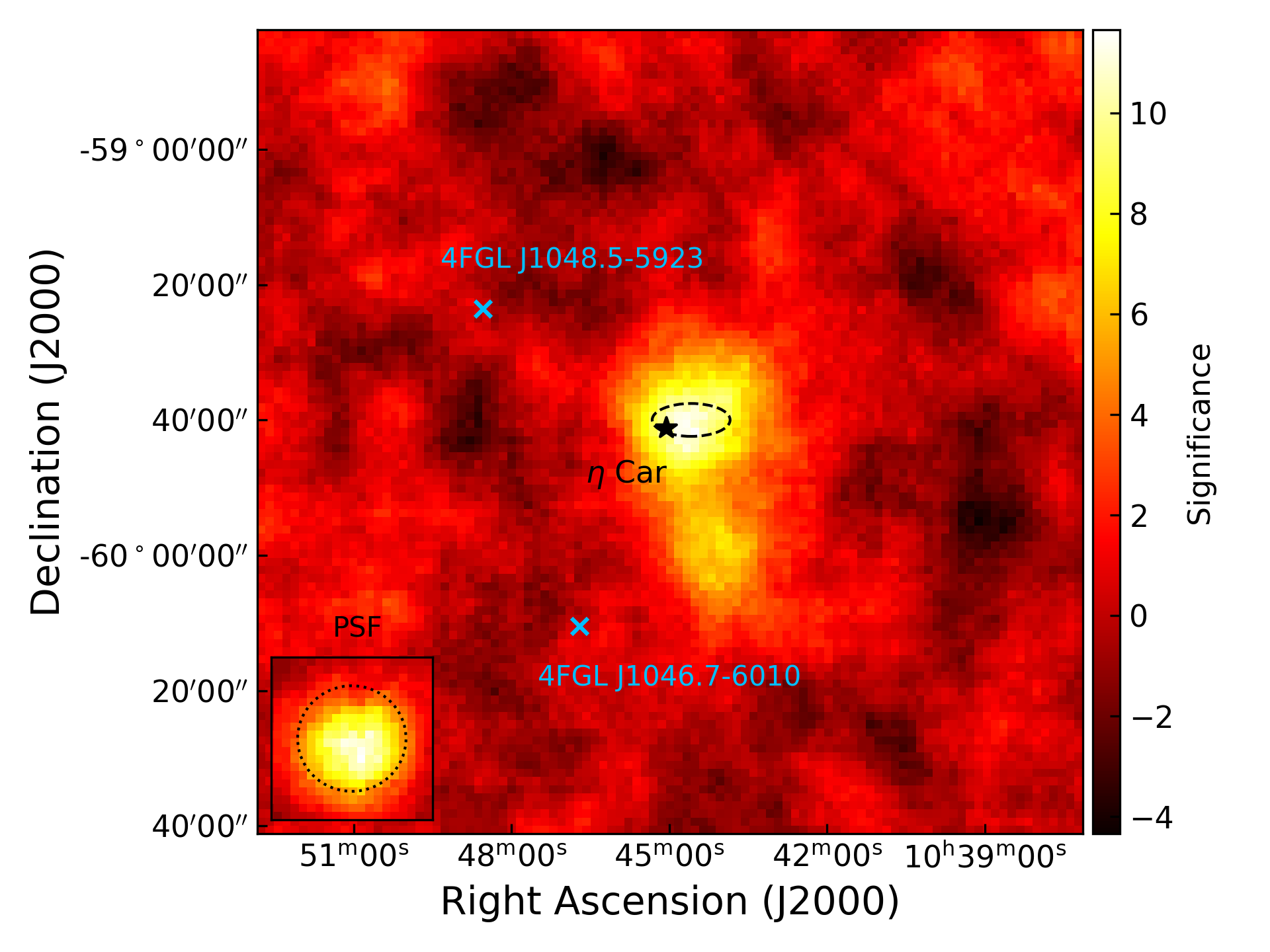}
\caption{\label{fig:MonoSign} Significance map as obtained
  in the CT5 mono analysis, for the combined DS-I ($0.78 < p < 0.96$)
  and DS-II ($1.09<p<1.10$) using an oversampling radius of
  0.1$^{\circ}$. The position of \EC\ is denoted with a black
  star and the long-dashed ellipse refers to the uncertainty in
  the positional fit as described in the main text. The positions of
  the nearby \fermi\ sources 4FGL~J1048.5$-$5923 and
  4FGL~J1046.7$-$6010 are shown with blue crosses
  \citep{Fermi4FGL}. The inlay shows the point spread function (PSF)
  as obtained using run-wise simulations (RWS) in colour, while the overlaid dotted circle
  indicates the PSF using classical simulations.}
\end{figure}

As a first step, we attempt to determine the morphology of the \g-ray
emission based on Fig.~\ref{fig:MonoSign} and the \g-ray maps of the
individual data sets. In ground-based \g-ray astronomy, instrument
response functions (IRFs) are typically produced using Monte-Carlo
(MC) simulations at fixed phase space points and assuming an ideal
detector (no turned-off pixels, uniform NSB, no telescope
tracking). The point spread function (PSF), however, is influenced by
the high NSB level: NSB photons can enter the reconstruction, which
leads to noisier shower images and hence an increased PSF. For complex
FoVs and measurements at the systematic limit, this approach is not
accurate enough anymore. It has been shown that a realistic
description of the observing conditions and instrument properties
under which \g-ray data has been taken is crucial when deriving the
PSF of \hess\ in measuring the extension of the Crab Nebula at TeV
energies \citep{AbdallaCrab2019}. With its challenging observing
conditions, the analysis presented here is another prime example for
the utility of this approach. In the following, we employ the usage of
run-wise simulations (RWS) to derive the PSF of DS-I and
DS-II. Although this more realistic description represents a
significant improvement over the classical IRFs, we note that due
to computational limitations, only the \g-ray signal is simulated
and not the cosmic-ray background, whose uniformity across the FoV
will be impacted by the increased and inhomogeneous NSB. A
non-uniformity in cosmic-ray background will impact the precision of
the background subtraction required to extract the \g-ray signal.

Positional fits of the \g-ray emission of DS-I and DS-II have
been performed using RWS assuming different underlying spectral indices
of the emission. Assuming a spectral index of $\Gamma=3.7$, consistent with
the spectral result discussed below, the morphology fit on DS-II (the
data set, which shows the lowest instrumental systematics) results in
a morphology consistent with being point-like and a best-fit
position of RA=\HMS{10}{44}{35}$\pm 6.6^{\mathrm{s}}$ and
Dec=\DMS{-59}{39}{56.6}$\pm 0.8'$ (J2000), $3.8'$ away from the optical
\EC\ position. To estimate the systematic uncertainty of the derived
best-fit position of data taken under different observing and
instrument conditions, morphology fits to DS-I and DS-II in the RWS
framework, and with classical IRFs in both the main and cross-check
analysis were performed. The derived best-fit position depends, for
instance, on the assumed underlying power-law spectral index, which
was varied between $\Gamma=3.0$ and $\Gamma=4.0$. The systematic error
on the position is estimated following this description as
$\Delta$RA=$\pm 1.5^{\mathrm{m}}$ and $\Delta$Dec=$\pm
4.8'$. Within the statistical and systematical error, the
derived best-fit position is consistent with the optical position of
\EC. The consistent treatment of the NSB by using RWS
is necessary to reconstruct a source that is compatible with being
point-like at the \EC\ position. 

Another complication in the determination of the source morphology is
the assumption of a single component as origin of the \g\
rays. Figure~\ref{fig:MonoSign} suggests that a weaker emission component
south of \EC\ exists and possibly biases the 2D morphology
fit. A dedicated analysis was performed towards this emission component at
RA=\HMS{10}{44}{22.8} and Dec=\DMS{-59}{57}{51.8} (J2000), which results
in a hotspot at 6.5$\,\sigma$ level in DS-II -- dubbed
HOTS~J1044$-$5957. At this significance level, and given the fact that the emission is located in
the region with the highest NSB in the FoV, we do not claim a new
source, but we note that not fully understood systematics could explain at
least parts of this emission. The contribution of HOTS~J1044$-$5957 to the
\EC\ emission is estimated to $\sim$15\%, based on a Gaussian fit
to the 1D spatial profile along the axis connecting \EC\ and
HOTS~J1044$-$5957. 

We emphasise that the hotspot is not detected in the cross-check
analysis, neither in DS-I nor in DS-II. This further supports that
the hotspot seen in the main analysis may be caused by a systematic
effect, likely the response of the classifying parameters to the
NSB. Also, at lower \g-ray energies, no counterpart is reported in
the \fermi\ 4FGL catalogue \citep{Fermi4FGL}. The spectral
properties of \EC\, as presented in the next section, are well
compatible within errors in the main and cross-check analysis,
implying a robust spectral measurement within uncertainties. To
conclude, all tests confirm the integrity of the signal, and given the
good positional coincidence, we assume in the following that it is
connected to the CWB system. The potential impact of the systematics
discussed above will be addressed as systematic error on the inferred
spectral measurement, as described in the next section.

\subsubsection{VHE \g-ray spectrum and light-curve}

\begin{figure}
\centering
\includegraphics[width=0.495\textwidth]{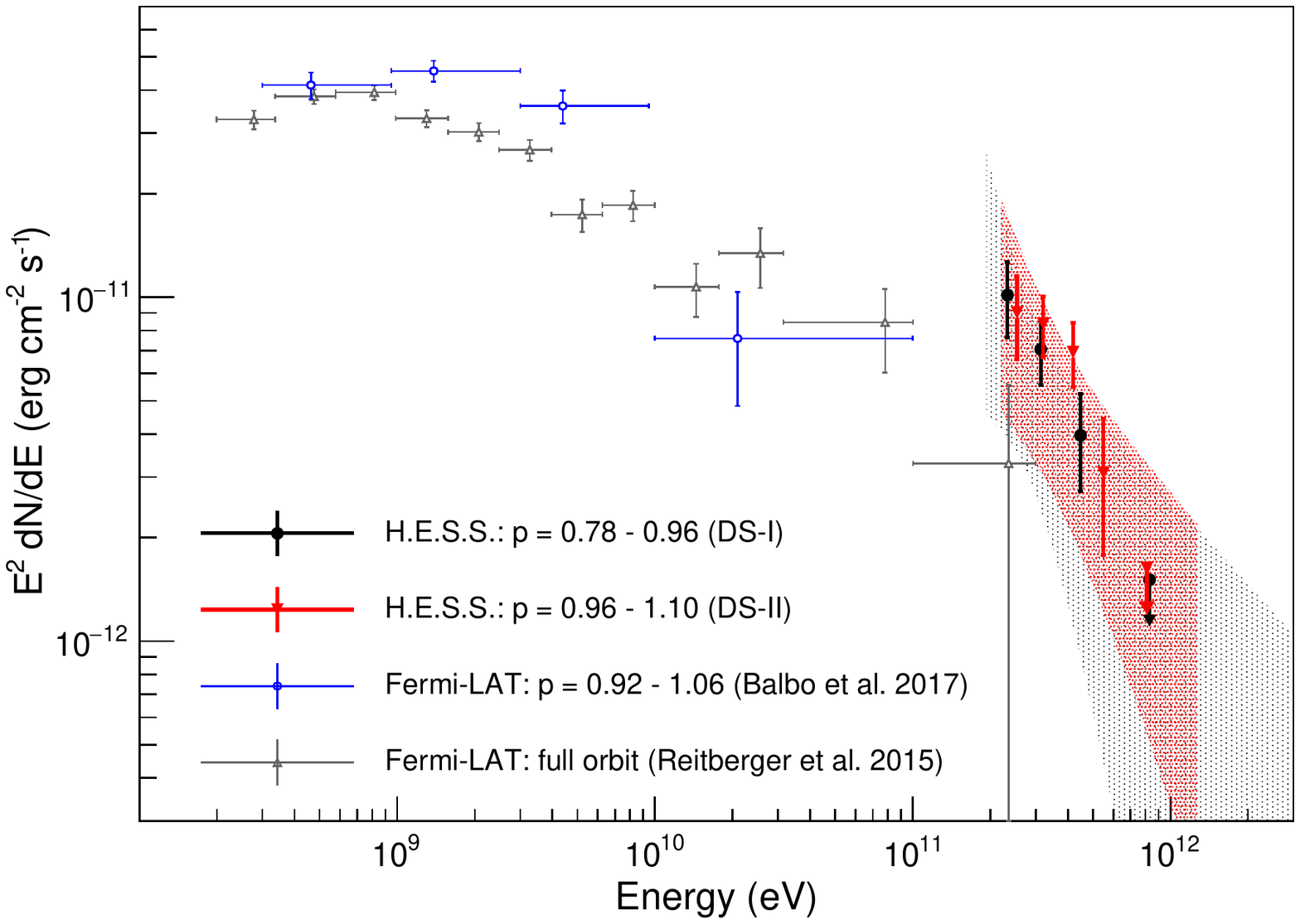}
\caption{\label{fig:Spec}Spectral energy distribution of \EC\ for DS-I (black)
  and DS-II (red). H.E.S.S. points show 1$\sigma$ statistical
  errors. The shaded regions indicate the combined statistical and systematic
  errors (as given in the main text and Tab.~\ref{tab:syst}). \fermi\ 
  spectra from \citet{Reitberger2015} for the full orbit (grey) and
  for the last periastron passage from \citet{Balbo2017} (blue) are
  also shown.}
\end{figure}

In the following, the intrinsic \g-ray spectrum is reconstructed
and the \g-ray signal is checked for variability and phase-locked
  flux variations as seen at other wavelengths. To enclose the entire
emission seen from \EC, the \textit{reflected background} method
\citep{Berge2007} was used with an on-source region radius of
0.2$^\circ$, and the forward-folding technique by \citet{Piron2001}
is applied to extract spectral parameters. RWS are currently
limited to the simulation of point sources of \g\ rays. For the
spectrum reconstruction one typically assumes that the response to
\g-ray like background events is identical in the signal and
background control regions. The number of turned-off pixels is much
higher towards \EC\ than in the surroundings. Hence, the response to
\g-ray like background events will also be different. For these
reasons, we use in the following classical IRFs.

Figure~\ref{fig:Spec} shows the spectra for DS-I and DS-II for the
mono analyses. The best-fit spectral index for DS-I is
$\Gamma_{\mathrm{DS-I}} = 3.94\pm 0.35_{\mathrm{stat}}$ above an energy
threshold of 190\,GeV. The flux normalisation at the decorrelation
energy \footnote{The decorrelation energy $E_0$ is defined as $E_0 =
  \exp{\frac{cov(F_0,\Gamma)}{F_0\Delta\Gamma^2}}\,\mathrm{GeV}$ \citep{Abdo2009},
  where $cov$ is the covariance error matrix.} of
$E_{\mathrm{0,DS-I}} = 290$\,GeV is $F_{0,\mathrm{DS-I}} = (5.1\pm 0.5_{\mathrm{stat}}
)\times
10^{-11}$\,ph\,cm$^{-2}$\,s$^{-1}$\,TeV$^{-1}$. The \g-ray
spectrum for DS-II above an energy threshold of 220$\,$GeV is best
described by a power-law with index of $\Gamma_{\mathrm{DS-II}} =
3.49\pm 0.23_{\mathrm{stat}}$ and normalisation $F_{0,\mathrm{DS-II}} = (3.2\pm 0.3_{\mathrm{stat}}) \times
10^{-11}$\,ph\,cm$^{-2}$\,s$^{-1}$\,TeV$^{-1}$ at a decorrelation
energy of $E_{0,\mathrm{DS-II}} = 360$\,GeV. The DS-II spectrum is
within statistical errors consistent with the DS-I spectrum (see Tab.~\ref{tab:spec}).

\begin{table*}
  \centering
  \caption{Spectral statistics for the two
      analyses. Uncertainties are statistical. $\mathrm{E_{th}}$
    denotes the threshold energy, $\mathrm{E_{0}}$ the decorrelation
    energy, and $\Phi_0$ the flux at the decorrelation energy.}
  \begin{tabular}{@{}ccccccc}
    \hline
    DS &$\mathrm{E_{th}}$ & $\Gamma$ & $\mathrm{E_{0}}$ & $\Phi_0$ & $\Phi(>0.2)$\,TeV\\ 
       & [TeV] & & [TeV]
       & $[\mathrm{cm}^{-2}\,\mathrm{s}^{-1}\,\mathrm{TeV}^{-1}]$
       & $[\mathrm{cm}^{-2}\,\mathrm{s}^{-1}]$\\ \hline
    I & 0.19 & 3.94 $\pm$ 0.35& 0.29 & (5.1 $\pm$ 0.5) $\cdot10^{-11}$ & (1.6 $\pm$ 0.2)  $10^{-11}$\\
    II & 0.22 & 3.49 $\pm$ 0.23 & 0.36 & (3.2 $\pm$ 0.3) $\cdot10^{-11}$ & (2.0 $\pm$ 0.2)  $10^{-11}$\\
  \end{tabular}
  \label{tab:spec}
\end{table*}

Several sources of uncertainty influence these spectral measurements. 
They range from MC simulations of the air-shower and the instrument,
the calibration, and data quality selection, to the high-level
reconstruction, \g-ray selection, and background
estimation. Table~\ref{tab:syst} lists all contributions to the flux
normalisation and spectral index uncertainty and is based on
\citet{2017AGNHessII}. The
systematic error from switched-off pixels has been increased from
the nominal 5\% to 10\% to reflect the conditions in this
analysis. Errors on the spectral index and normalisation from uncertainties
in the reconstruction and selection cuts are based on differences
between the main and cross-check analysis. The downward systematic
errors resulting from uncertainties in the background subtraction,
as discussed at the end of the previous section, were estimated from
the reconstructed integral flux of HOTS~J1044$-$5957 in DS-II. 
Furthermore, we note that the reconstructed spectrum of the lead
analysis may be prone to a shift of the energy scale of 20\% towards lower
energies. Even with this potential bias, the shifted spectral points
would still be captured by the systematic error budget estimate
presented in Tab.~\ref{tab:syst} and Fig~\ref{fig:Spec}. We
emphasise that the re-binned spectral points are shown only to guide
the readers eye and cannot easily be combined with \fermi\ data
points in a multiwavelength spectral fit.

\begin{table*}[ht]
  \caption{Estimated contributions to the systematic uncertainties in
    the spectral measurements for the mono analyses presented in this
    work follow the description in \citet{2017AGNHessII} and we assume
    similar systematics as for the PG~1553+113 source, which has a similarly
    steep spectral index as \EC. Values quoted are for upward /
    downward or symmetric errors. See text for more details on
    the differences. The energy scale uncertainty in this analysis is
    estimated to be larger than for PG~1553+113.}
\centering
\begin{tabular}{c c c c c c c}
\hline
Source of Uncertainty & Energy Scale & Flux & Index & Flux & Index\\
 & & DS-I & DS-I & DS-II  & DS-II \\
\hline
MC shower interactions			& & 1\%	&  & 1\%  & \\
MC atmosphere simulation		& 7\% & & & & \\
\hline
Instrument simulation / calibration	& 10\% & 10\% & & 10\% & \\
Broken pixels				& & 10\% & & 10\% & \\
Live Time				& & $<5\%$ & & $<5\%$ & \\
\hline
Reconstruction and selection cuts	& 15\%	& 15\%
                                            & 0.96 & 15\% & 0.34 \\
Background subtraction			& & 10\% / 55\% & 0.46 & 10\%
                                                                 / 55\% & 0.46 \\
\hline
Total					& 19\%	& 24\% / 60\%
                                            & 1.06  & 24\% / 60\% & 0.57 \\
\hline
\end{tabular}
\label{tab:syst}
\end{table*}

Figure~\ref{fig:LC} shows the \g-ray light-curve of \hess\ for the data
described in Tab.~\ref{tab:data}. All flux points with a
statistical significance of more than 2$\sigma$ are displayed as
points, whereas upper limits at 95\% confidence level are shown for
non-significant flux measurements. The \hess\ light-curve is
within uncertainties consistent with no variability\footnote{The
    probability for a constant flux in nightly and monthly bins,
    including fits to the deep-exposure data sets at p=0.95 and
    p=1.09, is at the 0.5\% level. When including systematic errors,
    this probability increases and the search for variability is
    non-significant at the $1.5 - 2.0$ $\sigma$ level.} This is in
agreement with the \fermi\ high-energy light-curve that is, however,
rather limited in statistics and can only probe the \g-ray emission on
much longer timescales. In the following, we will put the
\hess\ results in the multiwavelength context and discuss what they
imply for particle acceleration and \g-ray emission processes in this
unique object. 

\begin{figure}
\centering
\includegraphics[width=0.475\textwidth]{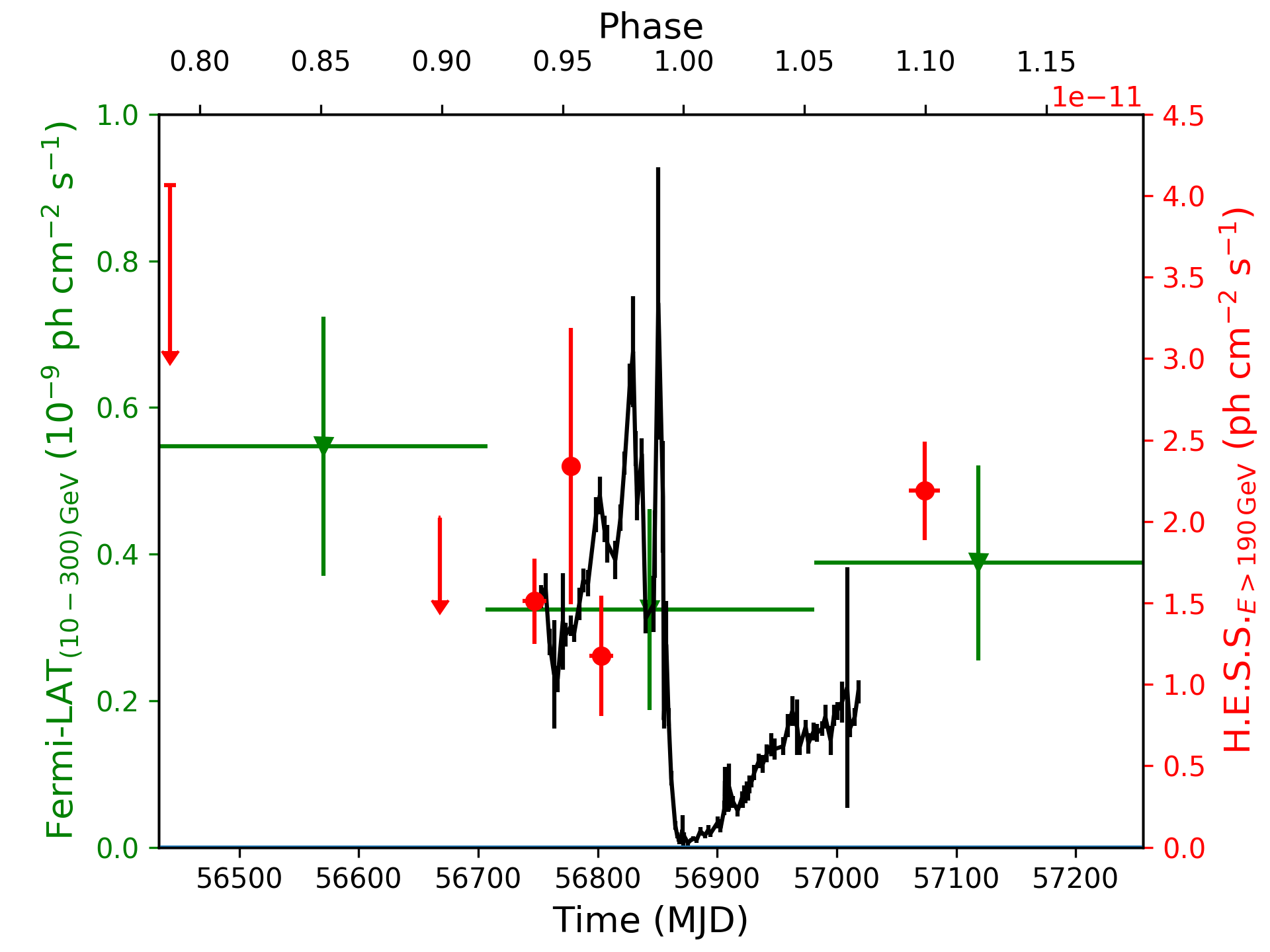}
\caption{\label{fig:LC} Phase-binned \hess\ flux above 190$\,$GeV as
  red points. The phase errors in the \hess\ light-curve
  points refer to the re-grouped data given in
  Tab.~\ref{tab:data}. \fermi\ \g-ray flux points from \citet{Balbo2017} and
  Swift X-ray data from \citet{Corcoran2017} are shown as green triangles
  and as black line, respectively. Flux errors are all 1$\sigma$
  while upper limits are 95\% confidence level. The X-ray data has
  been scaled to arbitrary units for better comparison to the \g-ray
  data.}
\end{figure}


\section{Discussion and outlook}
\label{sec:discussion}

The \hess\ measurement shows that there exists a VHE \g-ray source at a position coincident with the optical \EC\ position. The emission is consistent with being point-like using RWS and the \hess\ VHE \g-ray spectrum extends within statistical and systematic uncertainties at about 200\,GeV to the \fermi\ spectrum as presented in \citet{Balbo2017}. The \hess\ spectrum contains information about the maximum energy that the radiating particles in \EC\ can reach. The last bin in the \hess\ spectrum with a $>$$2\sigma$ statistical significance starts at $\sim$400\,GeV in the two data sets and was confirmed by both analysis chains.

While the 100\,MeV$ - $10\,GeV emission detected by \fermi\ can be interpreted either in a leptonic or a hadronic scenario, there seems to be a preference in the community that the $\geq$10\,GeV emission can be best explained as originating from interacting protons~\citep{Farnier2011,Bednarek2011,Ohm2015, Reitberger2015, Balbo2017}. In the following, we will elaborate on the implications of the maximum measured energy in the \hess\ spectrum in a leptonic and hadronic scenario.

Assuming that particles are accelerated in the CWR via the diffusive shock acceleration process, the time to accelerate a charged particle to an energy $E_{\mathrm{TeV}}$ in a magnetic field $B_G$ and in a shock with speed $v_{\mathrm{sh}, 10^3\mathrm{\,km\,s^{-1}}}$ can be estimated as \citep{Hinton2009}:

\begin{equation}
\tau_{\mathrm{acc}} = 5\times 10^4\,\eta\,v_{\mathrm{sh}, 10^3\mathrm{\,km\,s^{-1}}}^{-2}\,\frac{E_{\mathrm{TeV}}}{B_{G}}\,\mathrm{s},
\end{equation}
where $B_G$ is the magnetic field strength in Gauss, and $v_{\mathrm{sh},10^3\,\mathrm{km\,s}^{-1}}$ is the shock speed in 10$^3$\,km\,s$^{-1}$.
The $\eta$ parameter is the diffusion coefficient expressed in terms of the Bohm diffusion coefficient, $\eta = \kappa / \kappa_{\mathrm{Bohm}} \geq$1. In the most optimistic case, acceleration proceeds in the Bohm limit ($\eta=1$), and with the shock speed set by the faster, companion star wind: $v_{\mathrm{w}} = v_{\mathrm{sh}} = 3000\mathrm{\,km\,s^{-1}}$. 

Assuming that the radiating particles are electrons, VHE \g~rays are produced by inverse Compton scattering off the photon fields from the two stars. Given their spectral temperature, inverse Compton scattering is proceeding into the Klein-Nishina regime, where most of the electron energy is transferred to the \g~ray, at tens of GeV energies. Hence, \g-ray emission up to 400\,GeV implies maximum electron energies in the same range. If adiabatic losses are neglected \citep[c.f.][]{del-Palacio2016}, electrons at these energies predominantly suffer synchrotron and inverse Compton losses, while being accelerated. In the loss-limited case, $\tau_{\mathrm{acc}} = \tau_{\mathrm{cool}} = \tau_{\mathrm{IC+sync}}$ applies. Assuming negligible inverse Compton losses, the cooling time can be approximated as:

\begin{equation}
\tau_{\mathrm{cool}} = 4\times 10^2\,E_{\mathrm{TeV}}^{-1}\,B_{G}^{-2}\,\mathrm{s}.
\end{equation}

For electrons to reach an energy of 400\,GeV in the synchrotron-loss limited acceleration case (where $\tau_{\mathrm{acc}} = \tau_{\mathrm{sync}}$), the maximum allowed magnetic field is estimated to be $\sim$0.5\,$\eta^{-1}$\,G. With inverse Compton and adiabatic losses included, and a more realistic $\eta$ value, the maximum allowed magnetic-field strength is reduced even further.

In a hadronic scenario, protons and heavier nuclei would have to be accelerated to at least 5\,TeV \citep[e.g.][]{Hinton2009}. The maximum energy that protons can reach is to first order governed by the magnetic field strength and shock speed in the CWR for non-relativistic diffusive shock acceleration, and limited by the residence time of particles in the acceleration zone, as well as the particle density at the location of $p$-$p$ interaction. Typical densities in the CWR on the primary and companion side range between a few times 10$^{9}$ and $\sim$10$^{8}$ particles per cm$^{3}$, respectively, which implies $p$-$p$ cooling times of the order of $5-100$ days at phases considered here \citep[e.g.][]{Ohm2015}. According to Eq.~1, and for magnetic field strengths of the order $\sim$Gauss, proton energies $\geq$50\,TeV can be reached in the $p$-$p$ cooling-limited case and assuming Bohm diffusion. While the considerations made above might suggest a preference for a hadronic interpretation of the \hess\ emission, the present data situation in the VHE domain does not allow to draw a firm conclusion.

The measurement of variability, or phase-locked flux variations on timescales of days to weeks, could help to identify the region of \g-ray production inside the CWR. \hess\ observed \EC\ shortly before the thermal X-ray maximum at phase $p\sim 0.95$ and after the X-ray minimum and recovery at phase $p\sim1.1$. The flux \hess\ observed shows no indication of phase-locked flux variations. Due to the sporadic sampling and limited sensitivity of the measurement, no statement on variability on timescales shorter than months can be made. The lack of strong flux variations in the \hess\ light-curve before and after the thermal X-ray minimum is broadly consistent with the behaviour in hard X-rays \citep{Hamaguchi2018} and GeV \g~rays \citep[e.g.][]{Balbo2017}. No flare similar to the one detected with {\it AGILE} \citep{EtaCar:Agile} in 2008 is seen in the \hess\ data. Within statistical and systematic uncertainties, no change in the reconstructed source flux, nor a change in spectral index, could be detected. What seems apparent is that the steep spectrum observed with \hess\ is in tension with models that assume negligible \g\g\ absorption at $>100$\,GeV energies \citep[e.g][]{Gupta2017}. In fact, a strong phase-and energy-dependent suppression of the observed \g-ray emission around periastron passage is expected given the stellar parameters and the orbital configuration. Depending on the orbital solution \citep[][]{Madura2012}, a $\tau_{\gamma\gamma}$ of 1 - 10 can be estimated, and leads to a strong suppression of the \g-ray flux in the \hess\ energy range for the data presented here. Despite the comparably large systematics of the \hess\ measurement in this data set, the sensitivity of the presented observations demonstrates the potential of ground-based \g-ray observations for the study of short-timescale flux changes in systems like \EC\ (cf. Figs.~\ref{fig:Spec}, \ref{fig:LC}).

The \hess\ telescopes have been regularly monitoring \EC\ since 2014. In this paper, we present data taken until 2015. The remaining data together with planned observations of the upcoming periastron passage in early 2020 will then conclude one full orbit of \EC\ observations with \hess\ and be covered in a future publication. They will allow us to test for temporal flux changes along the orbit. These observations also cover the phase of maximum stellar separation, where the high-energy \g-ray flux measured by \fermi\ reaches its minimum. More importantly, the \hess\ observations of the upcoming periastron passage will be able to study the \g-ray emission on week-to-month timescales. This is crucial to probe the aforementioned X-ray maximum and minimum, which lasts between 30-60 days. The next-generation Cherenkov telescope array, CTA, will be able to improve considerably on the measured spectrum and light-curve. However, as demonstrated in this work, an accurate treatment of the increased NSB in the \EC\ region and careful treatment of the data with tailored simulations are essential.

\begin{acknowledgements}
  The support of the Namibian authorities and of the University of Namibia in facilitating 
the construction and operation of H.E.S.S. is gratefully acknowledged, as is the support 
by the German Ministry for Education and Research (BMBF), the Max Planck Society, the 
German Research Foundation (DFG), the Helmholtz Association, the Alexander von Humboldt Foundation, 
the French Ministry of Higher Education, Research and Innovation, the Centre National de la 
Recherche Scientifique (CNRS/IN2P3 and CNRS/INSU), the Commissariat à l’énergie atomique 
et aux énergies alternatives (CEA), the U.K. Science and Technology Facilities Council (STFC), 
the Knut and Alice Wallenberg Foundation, the National Science Centre, Poland grant no. 2016/22/M/ST9/00382, 
the South African Department of Science and Technology and National Research Foundation, the 
University of Namibia, the National Commission on Research, Science \& Technology of Namibia (NCRST), 
the Austrian Federal Ministry of Education, Science and Research and the Austrian Science Fund (FWF), 
the Australian Research Council (ARC), the Japan Society for the Promotion of Science and by the 
University of Amsterdam. We appreciate the excellent work of the technical support staff in Berlin, 
Zeuthen, Heidelberg, Palaiseau, Paris, Saclay, Tübingen and in Namibia in the construction and 
operation of the equipment. This work benefited from services provided by the \hess\
Virtual Organisation, supported by the national resource providers of
the EGI Federation. We thank Matteo Balbo, Klaus Reitberger as well as
  Mike Corcoran for providing us with multiwavelength data.
\end{acknowledgements}


\bibliographystyle{aa}
\bibliography{EtaCar_HESS2}

\end{document}